\documentclass{webofc}
\usepackage[varg]{txfonts}   
\begin{document}
\title{Heavy-quark dynamics in a hydrodynamically
evolving medium}
%
%

\author{\firstname{Marlene} \lastname{Nahrgang}\inst{1}\fnsep\thanks{\email{marlene.nahrgang@subatech.in2p3.fr}} \and
        \firstname{J\"org} \lastname{Aichelin}\inst{1} \and
        \firstname{Pol Bernard} \lastname{Gossiaux}\inst{1}\and
        \firstname{Klaus} \lastname{Werner}\inst{1}
}

\institute{ SUBATECH, UMR 6457, IMT Atlantique, Université de Nantes,
IN2P3/CNRS, 4 rue Alfred Kastler, 44307 Nantes cedex 3, France }

\abstract{In this talk we will discuss the recent advances in describing heavy-quark dynamics in the quark-gluon plasma (QGP), which evolves hydrodynamically. Special emphasis is put on the collective flow of the heavy-quarks with the medium constituents, for which we present our latest results obtained within the MC@sHQ+EPOS2 model at $\sqrt{s}=5$~TeV.
}
\maketitle
\section{Heavy quarks as probes of the QGP medium}
\label{intro}
Heavy quarks, especially charm and bottom, have since long been considered valuable probes for properties of the QGP. While the ``standard'' model of describing the space-time evolution of the bulk medium produced in heavy-ion collisions typically relies on a plasma phase that can be described by hydrodynamics, the heavy-quarks are produced in initial hard scatterings and therefore not equilibrated with the QGP at $\tau_0$, the initial time of hydrodynamics. For the soft and light sector the typical observables are related to collective phenomena on the hydrodynamical hypersurface, for which the memory of microscopic interactions is lost. For heavy quarks, however, some of this memory is kept and we can thus study their dynamics in order to learn about the underlying QCD force.

Since it became possible to produce a significant number of charm and then also bottom quarks at RHIC and the LHC, the theoretical description has produced a variety of models. Although some of them are limited to average energy loss calculations in a medium of average temperature, most approaches follow these general steps in their description:
\begin{itemize}
 \item \textbf{Initial production:} For the initial production one relies on the theoretical results for momentum spectra in proton-proton collisions, like the FONLL \cite{FONLL} formalism. This gives a successful comparison to experimental data for inclusive spectra. For more exclusive spectra, like initial correlations between the produced heavy quark-antiquark pair, one applies event generators, either Pythia \cite{Sjostrand:2006za} or those that couple NLO pQCD matrix elements with parton showers, like POWHEG or MC@NLO \cite{Frixione:2002ik,Frixione:2003ei}. For nuclear collisions, it seems appropriate to include additional cold nuclear matter effects, like shadowing of low-momentum production, via sets of nuclear parton distribution functions \cite{Eskola:2009uj}. One has to keep in mind that these fits come with relatively large uncertainties.
 \item \textbf{In-medium interaction between heavy-quarks and medium:} After the heavy-quark formation time or the equilibration time of the QGP medium, $\tau_0$, the heavy-quarks start interacting with the medium. Depending on the dynamical evolution equation, see below, the interaction is either described effectively via Fokker-Planck transport coefficients or from scattering cross sections with the medium constituents. For both approaches, the local temperature and the velocities are obtained from a hydrodynamical evolution. This hydrodynamical evolution might also be replaced by a full partonic microscopic transport model. In order to obtain reliable results it is important that the space-time evolution of the QGP is well tested against the ample experimental data available for the bulk observables. 
 \item \textbf{Hadronization:} Since comparison to experimental data occurs on the level of hadronic particles, it is necessary to perform a hadronization around the transition temperature of the confinement-deconfinement phase transition. There are basically two different mechanism applied: coalescence of a heavy quark with a light quark of the medium, which is most likely to happen at small momenta, or fragmentation of a heavy quark, predominantly happening at larger momenta. While high-momentum fragmentation is rather well constraint from comparisons to proton-proton data, the modeling of coalescence in particular contains many unkown non-perturbative effects, which can at best be modeled in an effective way. In the resonance recombination model a hadronization process stemming from the same underlying interaction as the heavy-quark medium interaction is realized \cite{He:2010vw, He:2011qa}.
 \item \textbf{and eventually final hadronic interactions:} For a long time, hadronic final interactions seemed not so relevant for final $D$ meson spectra as the hadronic cross sections were expected to be small. There is, however, growing awareness that around the pseudo-critical temperature interaction can also be strong on the hadronic side \cite{Tolos:2013kva,Ozvenchuk:2014rpa}. For entering the era of precision measurement, it will become important to have a good handle on the hadronic final interactions as well.
\end{itemize}

\section{Hydrodynamics and heavy quarks}
\label{sec-1}

The hydrodynamical description of the space-time evolution of the QGP has been very successful in reproducing various experimental data from $p_T$-spectra to flow harmonics - albeit with different combinations of transport coefficients and initial conditions. The relativistic viscous hydrodynamical model describes the evolution of energy-, momentum and charge density according to the conservation equations:
\begin{equation}
 \partial_\mu T^{\mu\nu} = 0 \quad \quad  \partial_\mu N^{\mu} = 0\, 
\end{equation}
with the energy-momentum tensor $T^{\mu\nu}= (e+p)u^\mu u^\nu - p g^{\mu\nu} - \Pi\Delta^{\mu\nu} + \pi^{\mu\nu}$ and $N^{\mu}=nu^\mu+j^\mu$, which include the ideal part and viscous corrections.
The first approach to couple particles to a hydrodynamical evolution is motivated by the non-relativistic Brownian motion and uses Fokker-Planck dynamics. It has the advantage that a knowledge about the nature of the (quasi-)particles of the QGP, or even the assumption of quasiparticles, is not needed and the interaction is encoded in three transport coefficients, the drag coefficient $A$ and the longitudinal and transverse momentum diffusion coefficients $B_{||}$ and $B_\perp$. These coefficients depend on the heavy-quark momentum and the medium temperature. The evolution equation for the heavy-quark distribution $f_Q$ is then
\begin{equation}
 \frac{\partial}{\partial t}f_Q(t,\vec{p})=\frac{\partial}{\partial{p^i}}\left( A^i(\vec{p})f_Q(t,\vec{p})+\frac{\partial}{\partial{p^j}}\left[B^{ij}(\vec{p})f_Q(t,\vec{p})\right]\right)\, .
\end{equation}
For a numerical solution of a finite number of heavy-quarks in the medium, the Fokker-Planck equation can be recast to the Langevin equation for individual particles
\begin{equation}
 \frac{{\rm d}}{{\rm d}t}\vec{p}=-\eta_D(p)\vec{p}+\vec{\xi}\quad{\rm with}\quad \langle\xi^i(t)\xi^j(t')\rangle=\kappa\delta^{ij}\delta(t-t')\, ,
\end{equation}
where the transport coefficients in the two equations are related to each other differently depending on whether a pre-point, mid-point or post-point prescription for the time discretization in the stochastic integral of the Langevin process is applied \cite{He:2013zua}.

In order to satisfy detailed balance and thus to describe a system, which in the long-time limit reaches thermal equilibrium \cite{Walton:1999dy,Moore:2004tg}, the transport coefficients need to fulfill the fluctuation-dissipation theorem, which gives $\eta_D=\frac{\kappa}{2m_QT}$ and for the spatial diffusion coefficient $D_s=\frac{T}{m_Q\eta_D}$.

The Fokker-Planck or Langevin equation is a second moment approximation of the Boltzmann equation for assumed small momentum transfer
\begin{equation}
 \frac{{\rm d}}{{\rm d}t}f_Q(t,\vec{x},\vec{p})={\cal C}[f_Q]\, ,
\end{equation}
with the collision integral
\begin{equation}
{\cal C}[f_Q]=\int{\rm d}\vec{k}[w(\vec{p}+\vec{k},\vec{k})f_Q(\vec{p}+\vec{k})-w(\vec{p},\vec{k})f_Q(\vec{p})]
\end{equation}
with a gain term (first term) and a loss term (second term). The transition probabilities are obtained from the underlying interaction model. 

In order to obtain similar results for the Fokker-Planck and the Boltzmann equation it is important to consistently calculate one transport coefficient from the same underlying interaction model and fix the other two from the Einstein relations. It is an interesting question if there is a general consistency between the two approaches and at which point they produce different results for heavy-quark dynamics \cite{Das:2015ana}.

\section{The heavy-quark diffusion coefficient}
\label{sec-2}
The spatial diffusion coefficient is an important quantity to compare different model approaches. It can be obtained with varying precision and accuracy from lattice QCD calculations, various interaction models and from a model-to-data analysis.
\subsection{$D_s$ from lattice QCD}
Typical temperatures reached in heavy-ion collisions are close
to $\Lambda_{\rm QCD}$ and thus within the nonperturbative
regime of QCD. A natural choice would therefore be to turn to lattice QCD calculations. Lattice
QCD calculations at finite T  are performed
in Euclidean space, which makes it very difficult to extract dynamical quantities. The relevant transport coefficients need be inferred from calculable quantities
on the lattice, such as the correlation
function of conserved currents. This means that integral equations of the following type with kernel $K$ need to be inverted
\begin{equation}
 G(\tau;T) =  \int_0^\infty \frac{{\rm d}\omega}{2\pi}\rho_E(\omega;T)K(\tau, \omega; T)
\end{equation}
Here, a large precision of the lattice QCD result and a well-motivated ansatz for the spectral function is needed.
One then obtains the transport coefficient from the slope of the spectral function $\rho_E$ at $\omega=0$. For the momentum diffusion
\begin{equation}
 \frac{\kappa}{T^3} = \lim_{\omega\to0}\frac{2T\rho_E(\omega; T)}{\omega}
\end{equation}
and the spatial diffusion is then $D_s=\frac{2T^2}{\kappa}$. 

The range of $D_s$ observed in lattice QCD calculations \cite{Ding:2012sp,Francis:2015daa,Banerjee:2011ra} is $D_s\sim (2-7) (2\pi T)$. The current uncertainties are still large and certain approximations, concerning quenched QCD, heavy quark vs. charm mass, and/or the missing continuum extrapolation, still limit the reliability of these results.

\subsection{$D_s$ from interaction models}
There are a variety of different interaction models being used in the study of heavy-quark dynamics in the QGP. A reasonable understanding includes pQCD (inspired models) at large momenta and some effective modeling of non-perturbative effects at lower momenta and temperatures. We will not go into the details of all of these models, but give only a short explanation. The reader may turn to the references for more information. 
\begin{itemize}
 \item \textbf{pQCD (inspired models):} MC@sHQ+EPOS2: collisional energy loss with running coupling $\alpha_s$ and one gluon-exchange approximation; radiative energy loss for intermediate momenta, where incoherent processes dominate, and an effective LPM reduction of the emission spectra, Boltzmann evolution coupled to a $3+1$d ideal fluid dynamical calculation from EPOS2 initial conditions \cite{Gossiaux:2008jv, Nahrgang:2013saa, Nahrgang:2013xaa, Nahrgang:2014vza, Nahrgang:2016lst}; BAMPS, similar energy loss models as MC@sHQ, but the QGP evolution and heavy quarks are both treated in a full Boltzmann approach \cite{Uphoff:2012gb,Fochler:2013epa}; Djordjevic et al.: collisional and radiative energy loss in a finite size medium with dynamical scattering centers, running coupling and magnetic masses, but no space-time dynamics of the medium \cite{Djordjevic:2003be,Djordjevic:2006tw, Djordjevic:2008iz, Djordjevic:2011dd}.
 \item \textbf{nonperturbative approaches:} AdS/CFT \cite{Gubser:2006nz, Horowitz:2015dta,Moerman:2016wpv}; PHSD: off-shell transport of light quasiparticles with masses and width obtained from fits to the lattice QCD equation of state, full Boltzmann propagation of heavy-quarks \cite{Song:2015sfa,Song:2015ykw}; POWLANG: Langevin dynamics with transport coefficients inspired from lattice QCD \cite{Beraudo:2014boa}; TAMU: thermodynamic T-matrix approach using the lattice QCD internal/free energy of a static $Q\bar{Q}$ pair as input, comprehensive sQGP approach for the equation of state, light quark \& gluon spectral functions, quarkonium correlators and heavy quark diffusion, resonance correlations in the T-matrix naturally lead to recombination near $T_c$ from the same underlying interactions \cite{vanHees:2007me, Riek:2010fk, He:2010vw, He:2011qa, He:2013zua, Liu:2016ysz}.
\end{itemize}

\subsection{$D_s$ from model-to-data analysis}
In most of the above models an explicit or implicit dependence on the parameters of the model is included, in both the description of the soft/light and the heavy sector. Bayesian analysis allows to tune several of these parameters simultaneously by a systematic comparison to the available data. This has been performed with the Duke model of heavy-quark Langevin propagation + UrQMD hadronic rescattering of the $D$ mesons including a parametrized spatial diffusion coefficient in the QGP phase \cite{Xu:2017obm}
\begin{equation}
 D_s(T,p)= \frac{1}{1+(\gamma^2p)^2}(D_s2\pi T)^{\rm lin}(T;\alpha,\beta) + \frac{(\gamma^2p)^2}{1+(\gamma^2p)^2}(D_s2\pi T)^{\rm pQCD}(T, p)\, ,
\end{equation}
with $(D_s2\pi T)^{\rm lin}(T;\alpha,\beta) = \alpha(1+\beta(T/T_c-1))$. 
This parametrization contains a linear part at small temperature and momenta and smoothly couples to the pQCD result at higher temperatures and momenta.
The background $2$d fluid dynamical evolution is tuned via Bayesian analysis to reproduce the available experimental data for the bulk observables, which in addition to some technical parameters gives the temperature dependence of the shear and bulk viscosities \cite{Bernhard:2016tnd}. For the heavy-quark analysis various observables from different beam energies and experimental collaborations are used.
The final result for the heavy-quark diffusion coefficient as a function of temperature and of momentum is obtained from a multi-step statistical analysis, which not only provides the best fit for $\alpha$, $\beta$ and $\gamma$, but also provides the confidence level and correlations between these parameters. The obtained diffusion coefficient lies in the range of lattice QCD results and is at the lower end of the range covered by models. Please refer to \cite{Xu:2017obm} for details.

\section{Collective flow of heavy-quarks}
\label{sec-3}
In the remainder of this work we will discuss the possible origins of collective flow of the heavy quarks. For this we will show latest results obtained within the MC@sHQ+EPOS2 model for the highest LHC energy at $\sqrt{s}=5$~TeV in comparison to the (preliminary) experimental data from the CMS collaboration \cite{CMS:2016jtu}. Previously, we had published the first predictions for triangular flow of heavy quarks at $\sqrt{s}=200$~GeV and $\sqrt{s}=2.76$~TeV in \cite{Nahrgang:2014vza}. We compare two different models of energy loss, the purely collisional and the collisional plus radiative energy loss. Both include an overall parameter $K$, which is fitted to reproduce central $\sqrt{s}=2.76$ TeV data for the nuclear modification factor $R_{\rm AA}$ in the intermediate $p_T$ range. For this we find $K=1.5$ for the purely collisional case and $K=0.8$ for the collisional + radiative scenario.

\begin{figure}[t]
 \centering
 \includegraphics[width=0.99\textwidth]{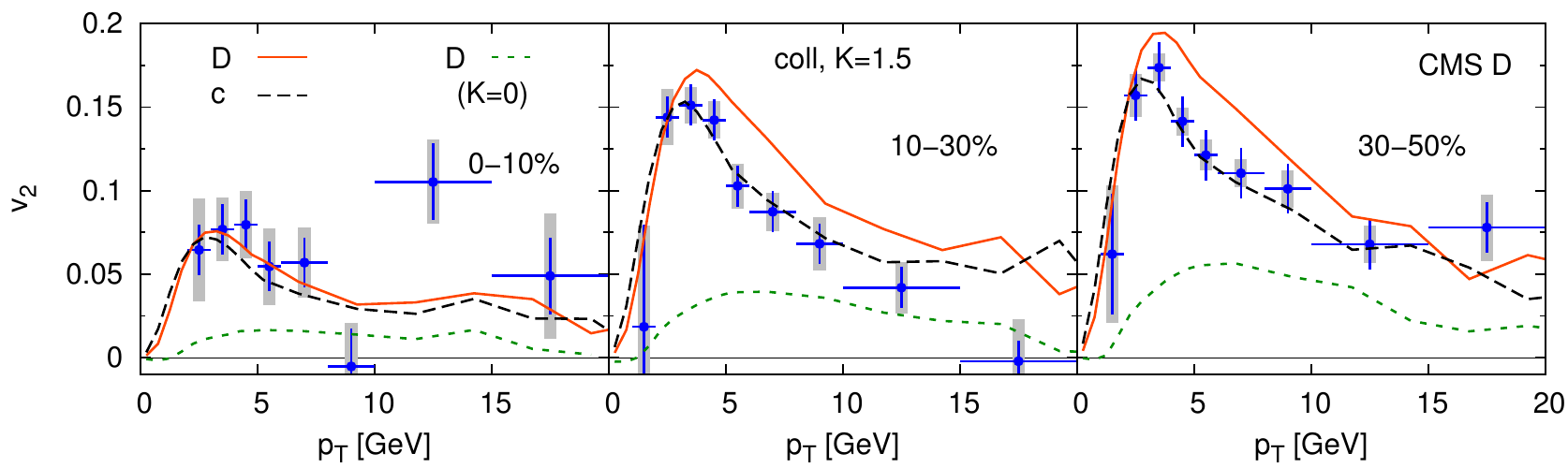}
 \caption{The centrality dependence of the elliptic flow of $D$ mesons and charm quarks as a function of transverse momentum for a purely collisional energy loss scenario, scaled by $K=1.5$. The solid (orange) curve is for final $D$ mesons, the long dashed (black) curve is for charm quarks at the hadronization temperature of $T_c=155$~MeV and the short dashed (green) curve is a scenario without any in-medium interactions (corresponding to $K=0$), where only coalescence contributes to the $D$ meson flow. Experimental data is for $D_0$ mesons from the CMS collaboration \cite{CMS:2016jtu}.}
 \label{fig:v2coll}
\end{figure}
\begin{figure}
 \centering
 \includegraphics[width=0.99\textwidth]{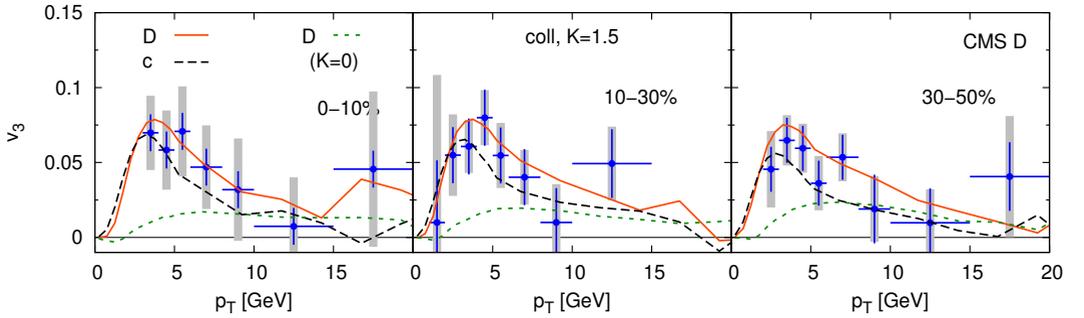}
 \caption{Same as Fig. \ref{fig:v2coll} but for the triangular flow.}
 \label{fig:v3coll}
\end{figure}

\begin{figure}
 \centering
 \includegraphics[width=0.99\textwidth]{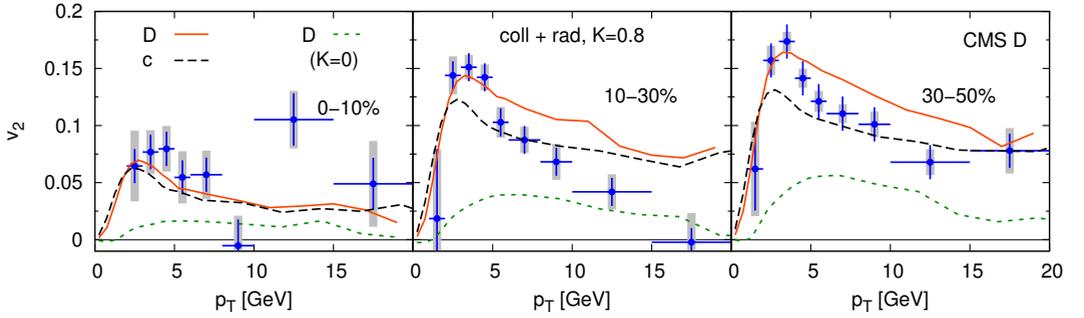}
 \caption{Same as Fig. \ref{fig:v2coll} but for the collisional + radiative energy loss mechanism with $K=0.8$.}
 \label{fig:v2collrad}
\end{figure}

\begin{figure}
 \centering
 \includegraphics[width=0.99\textwidth]{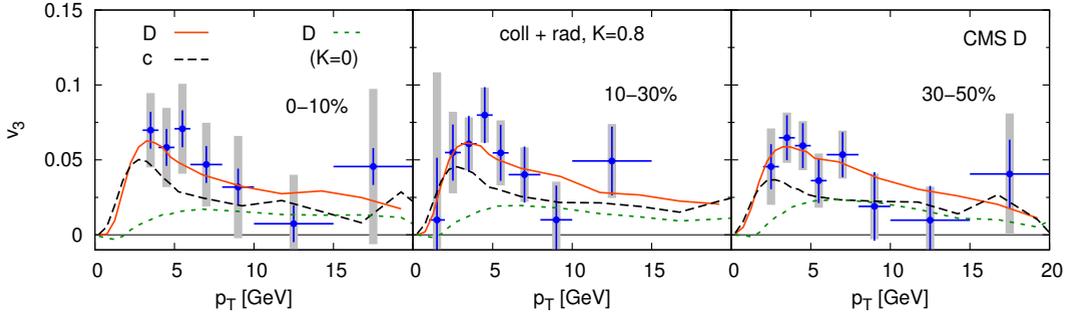}
 \caption{Same as Fig. \ref{fig:v2collrad} but for the triangular flow.}
 \label{fig:v3collrad}
\end{figure}

In Figs \ref{fig:v2coll} - \ref{fig:v3collrad} we compare the respective flow harmonics of $D$ mesons to the ones of charm quarks. During hadronization via coalescence, the charm quarks pick up flow from the light quark they recombine with. Therefore the $D$ meson $v_n$ is larger and somewhat shifted in $p_T$. Another way to look at this contribution is to switch off all in-medium interaction of the charm quarks in the QGP, which means their flow vanishes. The corresponding $D$ meson flow is then only stemming from coalescence, since hadronic interactions are not included in our model.

In the case of elliptic flow we see a clear centrality dependence of both the in-medium and the coalescence contributions, which is clear from the changing geometry. The origin of the triangular flow is, however, due to fluctuations in each centrality class and already the charged particle $v_3$ shows a very small centrality dependence. Since the medium size and temperature decrease with increasing centrality, it is expected that the charm flow picks up less and less flow from the medium in more peripheral collisions. We can observe in Fig. \ref{fig:v3coll} and \ref{fig:v3collrad} that the maximum of the charm quark flow is indeed decreasing for larger centralities. The final $D$ meson flow, is however the same, because the centrality dependence of the light parton flow.

In comparison to the experimental data we conclude that the tuned purely collisional energy loss mechanism seems to be favored. This is a consistent conclusion, which is also obtained from high-momentum $R_{\rm AA}$. It can be explained by the missing ingredients in the radiative contribution to the energy loss at higher $p_T$ in our model, which has its strengths in the intermediate $p_T$ range. At high momentum finite path length effects lead to additional coherent suppression of energy loss. Moreover, the current radiative vertex has fixed coupling of $\alpha_s=0.3$. Work in both of these direction is in progress.

\begin{figure}
 \centering
 \sidecaption
 \includegraphics[width=0.5\textwidth]{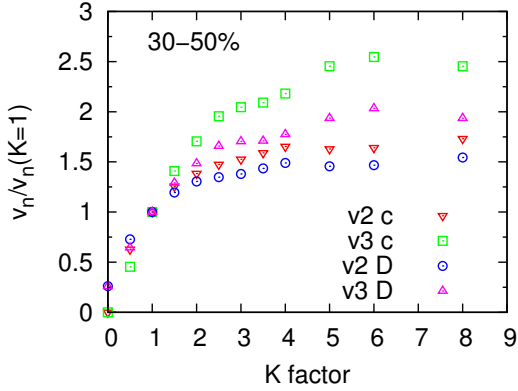}
 \caption{The dependence of the integrated flow harmonics $v_2$ and $v_3$ of charm quarks and $D$ mesons on the diffusion coefficient, tuned by a scaling factor $K$, which multiplies the scattering cross sections. Here, we focus on a purely collisional energy loss mechanism. The curves are scaled to their respective values at $K=1$ to visualize the relative spread.}
 \label{fig:Kdep}
\end{figure}

Finally, we investigate the sensitivity of the $v_2$ and the $v_3$ on the underlying diffusion coefficient, which scales directly with $K$. For different values of $K$ we calculate the integrated $v_2$ and $v_3$ for the collisional energy loss scenario in the $30-50$\% centrality class. The integrated flow harmonics are dominated by the abundant low-momentum charm quarks, where collisional energy loss is expected to the main mechanism. As expected Fig. \ref{fig:Kdep} shows that with increasing $K$ the charm quarks pick up more and more flow from the medium, while the coalescence contribution is independent of $K$ and therefore the $D$ meson flow does not show the same strong increase. Since the charm quarks can, however, not pick up more flow than is carried by the underlying medium expansion, one observes a saturation toward larger values of $K$. We normalize the curves by their respective values at the nominal $K=1$ in order to compare the relative sensitivity on $K$ and see that the higher-order triangular flow is more sensitive than the elliptic flow. We conclude that measuring the $v_3$ with high precision provides very helpful additional information in constraining the heavy-quark diffusion coefficient.

\section{Summary}
We have reviewed the general approaches to describe the heavy-quark dynamics by a coupling of either the Fokker-Planck or the Boltzmann equation with a (mostly) hydrodynamical evolution of the underlying QGP medium. The input transport coefficients for these calculations are obtained either from lattice QCD or from an appropriate interaction model or from a model-to-data Bayesian analysis.

Finally, we presented result for the elliptic and triangular flow harmonics of charm quarks and $D$ mesons, obtained in our model MC@sHQ+EPOS2. We showed the different contributions stemming from the in-medium interaction of charm quarks with the QGP (quasi)particles and during hadronization via coalescence. The relatively flat centrality dependence of the $D$ meson $v_3$ as seen in the experimental data, can be explained by a slightly decreasing charm flow and a slightly increasing component from coalescence. The relative sensitivity of $v_3$ on the transport coefficient is larger than that of the standard $v_2$. Measurements of $v_3$ with good precision are therefore valuable additional sources of information about the QCD interaction between heavy flavor and the QGP.

\end{document}